\documentclass[12pt]{article}
\usepackage{graphicx}
\usepackage{setspace}
\usepackage[super,sort&compress]{natbib}
\usepackage{multirow}
\begin{document}

\begin{center}
{\bf \large Magnetic Charge Transport}
\vspace{1cm}

{ S. T. Bramwell $^{1\ast}$, S. R. Giblin$^{2\ast}$, S. Calder$^1$, R. Aldus$^1$, D. Prabhakaran$^3$ and T. Fennell$^4$}
\vspace{0.5 cm}

{\small
{\it 1. London Centre for Nanotechnology and Department of Physics and Astronomy, University College London, 17-19 Gordon Street, London, WC1H OAH, U.K.}
\vspace{0.5 cm}

{\it 2. ISIS Facility, Rutherford Appleton Laboratory, Chilton, Oxfordshire OX11 0QX, U.K.}
\vspace{0.5 cm}

{\it 3. Clarendon Laboratory, Department of Physics, Oxford University, Oxford, OX1
3PU, U.K. }
\vspace{0.5cm}

{\it 4. Institut Laue-Langevin, 6 rue Jules Horowitz, 38042 Grenoble, France.}
}
\end{center}

{\bf 
It has recently been predicted that certain magnetic materials contain mobile magnetic charges or `monopoles'~\cite{CMS}. Here we address the question of whether these magnetic charges and their associated currents (`magnetricity') can be directly measured in experiment, without recourse to any material-specific theory. By mapping the problem onto Onsager's theory of weak electrolytes, we show that this is possible, and devise an appropriate method. Then, using muon spin rotation as a convenient local probe, 
we apply the method to a real material: the spin ice Dy$_2$Ti$_2$O$_7$. Our experimental measurements prove that magnetic charges exist in this material, interact via a Coulomb interaction, and have measurable currents. We further characterise deviations from Ohm's Law, and determine the elementary unit of magnetic charge to be 5 $\mu_B{\rm \AA}^{-1}$, which is equal to that predicted by Castelnovo, Moessner and Sondhi~\cite{CMS} using the microscopic theory of spin ice. Our demonstration of magnetic charge transport has both conceptual and technological implications.}
\newpage

\section{Introduction}
The transport of electrically charged quasiparticles (based on electrons, holes or ions) plays a pivotal role in modern technology as well as determining the essential function of biological organisms.  In contrast, the transport of magnetic charges has barely been explored experimentally, mainly because magnetic charges are generally considered to be, at most, convenient macroscopic parameters~\cite{Morrish} rather than sharply defined quasiparticles. However, the recent proposition of emergent magnetic monopoles in spin ice materials~\cite{CMS} may change this point of view. It is of great interest to investigate the transport of such atomic-scale magnetic charges, as the resulting `magnetricity' could have potential technological significance.  Furthermore, the direct observation of magnetic charge and current would afford the strongest case for believing in their reality. Thus, in the analogous case of electrical charge transport (electricity), the fact that both current $I$ and the elementary charge $e$ can be directly measured emphasises that these quantities are `real', rather than just convenient parameters in a model. 

Spin ices are frustrated magnets, a class of magnetic material that is well known for supporting exotic excitations, both in experiment~\cite{Broholm} and in theory~\cite{MoessnerChalker,balents}. The spin ice family~\cite{Harris,Ramirez,Bram,Schiffer} is predicted to support sharply defined magnetic monopole excitations and offers the possibility of exploring the general properties of magnetic charge transport in an ideal model system~\cite{CMS}. There is significant experimental evidence to support the existence of spin ice magnetic monopoles~\cite{CMS, Jaubert}, but this evidence does not include the direct observation of charge or current and relies strongly on interpretation via the microscopic theory of spin ice. In the analogous electrical case, charge and current may be directly measured, with only the basic knowledge that the sample is say a semiconductor or a metal. It therefore seems realistic to seek a method of measuring magnetic charge and current that is similarly direct and robust.

As spin ices are magnetic analogues of water ice, the magnetic charges of spin ice are predicted to be directly analogous to water ice's mobile ionic defects. Water ice in turn may be classified as a  
weakly dissociated electrolyte, for which there exist a variety of general experimental methods to prove the existence of unbound ions and to estimate their charge and mobility. 
Although most of the relevant methods have no possible magnetic analogy, we have identified one that does. The `second Wien effect' describes the nonlinear increase in dissociation constant $K$ (or equivalently the conductance) of a weak electrolyte (solid or liquid) in an applied electric field $E$~\cite{Moore,Onsager}.  In a seminal work of 1934 Onsager~\cite{Onsager} derived a general equation for the second Wien effect. This equation provides an excellent description of experimental conductivity measurements, and remarkably for a thermodynamic relation, enables the determination of the elementary charge $e$.  

The aim of the current Article is to present a theory for the direct characteristion of magnetic currents via the Wien effect, and to test it experimentally. The principal theoretical predictions are that the magnetic conductivity is proportional to the fluctuation rate $\nu$ of the magnetic moment, and that the `elementary' magnetic charge $Q$ (which should depend on material) may be derived from the initial slope and intercept of the field dependence of $\nu(B)$ via the equation:
\begin{equation}\label{Q}
\tilde{Q} = 2.1223~m^{1/3}T^{2/3}.
\end{equation} 
Here $m = {\rm slope}/{\rm intercept}$ is the field gradient of the relative magnetic conductivity, $T$ is the temperature and the 
tilde means that $Q$ is measured in 
units of $~\mu_B{\rm \AA}^{-1}$ (SI units are used elsewhere). Eqn. \ref{Q} is valid for a sufficiently weak magnetic electrolyte and assumes a small relative permeability of the material (although this assumption is not strictly necessary: see below). 

Thus we predict a specific data collapse of $\nu(B,T)$ over the range of field and temperature where the theory is valid.  As the charge depends only on the ratio $\nu(B)/\nu(0)$, we simply require an experimental quantity that is {\it proportional} to $\nu(B)$ in the weak field limit. This use of relative quantities obviates the need for absolute measurements, which makes the method particularly robust and flexible. In the case of the spin ice Dy$_2$Ti$_2$O$_7$~\cite{Ramirez,Schiffer}, we have found that transverse field muon spin rotation~\cite{Uemura} affords a convenient probe. We have therefore used this technique to test the theory and to determine the elementary magnetic charge in the spin ice Dy$_2$Ti$_2$O$_7$. 

The Article is organised as follows. In the ``Theory of the Magnetic Wien Effect'' Section, we describe the magnetic equivalent of the second Wien effect, and derive Eqn. \ref{Q}.
In the ``Application to a Real Material'' Section we  describe our muon spin rotation ($\mu$SR) experiments on Dy$_2$Ti$_2$O$_7$. Finally, in the ``Discussion and Conclusions'' Section, we  summarise our main findings and discuss the broader implications of the work. 

\section{Theory of the Magnetic Wien Effect}\label{Two}
The dissociation of a weak electrolyte may generally be represented by two successive equillibria, the first representing the formation of an associated or bound ion pair and the second the dissociation into free ions. For example, in the specific case of water ice, the equilibria may be represented: 
\begin{equation}\label{ice}
{\rm 2H_2O = [H_3O^+OH^-] = H_3O^+ + OH^-},
\end{equation}
where $[\dots]$ represents the bound pair. Here the unbound ions interact according to a (possibly screened) Coulomb law. 
A physical picture of the Wien effect is that an applied field accelerates the free ions and, opposed by Brownian motion, in some cases does enough work to overcome the Coulomb potential barrier that binds the ions together. The result is an increase, with field, of the rate of dissociation and hence of the corresponding equilibrium constant. The field acts only on the
forward reaction of the second (dissociation) equilibrium via the electrical force $F = -eE$.   
Onsager's theory~\cite{Onsager} is valid under the condition that the concentration of unbound defects is sufficiently small for the Debye screening length to be much greater than the association distance~\cite{Onsager,Bass}.  His main result may be approximated by the following form in the weak field limit:
\begin{equation}\label{weak}
K(E) = K(0) \left( 1 + b + \frac{b^2}{3} \dots\right),
\end{equation}
valid for $b < 3$, where $b$ is the dimensionless group
\begin{equation}
b = \frac{e^3 E}{8\pi\epsilon k^2 T^2}.
\end{equation}
Here the symbols have their usual meaning and $\epsilon$ is the permittivity of the solvent.  
The quantity $-kT \sqrt{8b}$ may be interpreted as the Coulombic barrier to ion pair dissociation, at which the field energy $-eEr$ balances the force of Coulomb attraction~ \cite{Bass}  (note that here the zero of potential energy is measured at infinite separation).   

Onsager's theory would be expected to be very general and realistic. It should be applicable, by analogy, to emergent magnetic charges and their transport. In analogy with Eqn. \ref{ice}, the formation of magnetic charges may be represented by the successive equilibria: 
\begin{equation}
{\rm [solvent] = [bound~charges] = [free~charges]}
\end{equation}
With the correspondence $e \rightarrow Q, E \rightarrow B, \epsilon_0 \rightarrow \mu_0^{-1}$, we find that
the dissociation equilibrium into free magnetic monopoles is then characterised by Eqn. \ref{weak} with 
\begin{equation}\label{b}
b =  \frac{ \mu_0 Q^3 B}{8\pi k^2 T^2 \mu_r}
\end{equation}
where $\mu_r$ is the relative permeability of the magnetic `solvent'. A pictorial representation of the physical content of the theory is shown in Fig. 1. 

We define $n_b, n_u$ as the number of bound and unbound pairs respectively, $n_0 = n_b+n_u$ as the total pair concentration and  $\alpha = n_u/n_0$ as the degree of dissociation. 
The dissociation constant is given by
\begin{equation}
K = n_0 \frac{\alpha^2}{1-\alpha}.
\end{equation}
The increase in this equilibrium constant with increasing magnetic field (the Wien effect) defines a corresponding change in magnetic moment per unit forward reaction: 
\begin{equation}
\left(\frac{\partial \ln K}{\partial B}\right)_{T,N} = \left(\frac{\partial \Delta G^0/kT}{\partial B}\right)_{T,N} = \frac{\Delta \mu}{kT}, 
\end{equation}
where $\Delta G^0$ is the change in Gibbs energy per atom.
Using Eqn. \ref{weak} we find, for the weak field limit,
\begin{equation}\label{mag}
\Delta \mu =  \frac{k T b}{B},
\end{equation}
which is field-independent. 

Following Onsager we now assume that the reaction rates of the first equilibrium are much faster than those of the second equilibrium, in which case all molecules may be considered as bound pairs~\cite{Onsager}. Recalling that $\alpha \ll 1$,  we see that $n_0 \approx n_b \approx N$, so the total number of defects is approximately constant at a given temperature. Following a small disturbance, the relaxation of $\Delta \alpha$ back to its equilibrium value is determined by charge recombination. Onsager showed that the decay is exponential with time constant 
$
\nu = 2 \mu_0 \kappa. 
$
where $\kappa$ is the conductivity, which is proportional to the equilibrium $\alpha$. Following Eqn. \ref{mag} a fluctuation in $\alpha$ causes a proportionate fluctuation in magnetic moment, which decays at the same rate $\nu$. Thus, measurement of the magnetic moment fluctuation rate as a function of field is equivalent to the observation of the magnetic conductivity, and gives direct access to the Wein effect: 
\begin{equation}\label{final}
\frac{\nu(B)}{\nu(0)} = \frac{\kappa(B)}{\kappa(0)} = \frac{\alpha(B)}{\alpha(0)} =  \sqrt{\frac{K(B)}{K(0)}} = 1 + \frac{b}{2}+\frac{b^2}{24} + \dots,
\end{equation}
where $b$ is linear in field (eqn. \ref{b}). This equation should be valid provided that both $\alpha$ and its change in field $\Delta \alpha$ are sufficiently small. It may be used to derive Eqn. \ref{Q} in the case that $\mu_r =1$.

\section{Application to a Real Material}

In spin ice materials like Ho$_2$Ti$_2$O$_7$ and Dy$_2$Ti$_2$O$_7$, the magnetic charges or monopoles are predicted to be a consequence of the many body nature~\cite{Gingras} of the dipole-dipole interactions in these materials~\cite{CMS}. In 
detail, the Ising-like Ho or Dy  moments (`spins') are equivalent to proton displacement vectors in water ice~\cite{Bram}. The spins populate the vertices of a lattice of linked tetrahedra, pointing either `in' or `out' of any particular tetrahedron. The configuration `two spins in, two out' corresponds to a water molecule, H$_2$O. At low temperature, magnetic interactions of mainly dipolar character ensure a ground state consisting only of such `two in, two out' configurations: the magnetic system is governed by an ice rule and shares with water ice the Pauling zero point entropy~\cite{Ramirez}. 
The elementary excitation out of the spin ice state is a single spin flip, which 
breaks the ice rule on neighbouring tetrahedra to create a pair of 
defects, `three in one out plus one in three out'.  The equivalent water ice 
configuration is a closely bound ion pair H$_3$O$^+$OH$^-$~\cite{BramHarris,Ryz}. Castelnovo {\it et al.}~\cite{CMS} predicted that defects should unbind and behave as free magnetic monopoles (in the H-field) that interact via the magnetic Coulomb law. This interaction is determined only by fundamental constants except that the charge takes the value $\pm Q_{\rm theoretical} = 2\mu/a$, where $\mu \approx  10 \mu_B$ is the rare earth moment and $a_d = 4.3356  {\rm \AA}$ is the distance between tetrahedron centres. The formation of magnetic charges in spin ice is thus directly analogous to Eqn. \ref{ice}.  As water ice shows a Wien effect~\cite{Eigen}, it seems valid to test for one in spin ice.

We studied Dy$_2$Ti$_2$O$_7$ by the technique of transverse field muon spin rotation, with the aim of testing the theory of Section \ref{Two}, and of measuring the `elementary' charge $Q$. 
In $\mu$SR, muons implanted into a sample precess around the sum of the local and applied fields, and their decay characteristics give information on the time dependence of these fields. 
In an applied transverse field the muon relaxation function has an oscillatory form, resulting from the uniform muon precession about the applied field. However this uniform precession can be dephased by fluctuating local fields that arise from the sample magnetization ${\bf M}(r,t)$. In the low temperature limit of slowly fluctuating magnetization the dephasing may lead to an exponential decay envelope of the muon relaxation function in which the decay rate $\lambda$ depends only on the characteristic rate $\nu$ of the magnetic fluctuations. By dimensional analysis $\lambda \propto \nu$, so the key property $\nu(B)/\nu(0) = \lambda(B)/\lambda(0)$ can be directly measured (see Eqn. \ref{final}). In the opposite (high temperature) limit of fast fluctuations, $\lambda$ also depends on the width of the field distribution $\sigma$ and the muon gyromagnetic ratio $\gamma$. Here, one expects $\lambda \propto (\sigma \gamma)^2/\nu$ (Ref.~\cite{Uemura}), with $\sigma$ independent of temperature, so $\lambda$ is proportional to the relaxation timescale $\tau = \nu^{-1}$. 

The muon relaxation rate $\lambda(B)$ was measured as a function of field and temperature after zero field cooling of the sample to 60 mK (see Methods). Selected experimental results through the region of interest are shown in Fig. \ref{fig2}. 
To compare with theory it is useful to define an effective magnetic charge
by $\tilde{Q}_{\rm eff} \equiv 2.1223~m_{\rm experimental}^{1/3}T^{2/3}$
(see Eqn. \ref{Q}), clearly demonstrating that only {\it relative} changes in $\lambda$ are required to obtain an {\it absolute} measurement of $\tilde{Q}_{\rm eff}$.
Here we set $\mu_r = 1$, which is appropriate to spin ice under the conditions of interest (see Methods). 

We would expect $\tilde{Q}_{\rm eff}$ to approximate the true $Q$ only in a finite range of temperature. At too low temperature (say $b= 2 m B  \gg 3$) the theory breaks down because $\Delta \alpha$ becomes large, while at too high temperature, it breaks down because $\alpha$ becomes large.
These limits on the range of validity can be estimated experimentally. 

We first consider the temperature dependence of $\lambda(B=2 mT)$, as shown in Fig. \ref{fig3}. The high temperature behaviour mirrors the known behaviour of the magnetic relaxation time, decreasing with temperature as $e^{{\rm constant}/T}$ (Ref. ~\cite{Orendac,Jaubert}). At low temperature $\lambda$ increases with temperature, consistent with the expected $\lambda \propto \nu \propto \alpha$. From the graph we can determine that the crossover from the presumed unscreened regime to the screened regime with rapidly increasing $\alpha$ is at $T_{\rm upper}  \approx  0.3$ K. 
The inset of Fig. \ref{fig3} shows the temperature dependence of $\lambda(B=2 mT, 1 mT)$: the apparent irregularity of the measurements below $T < T_{\rm upper}$ is in fact consistent with the data collapse predicted by theory (see below), and naively demonstrates that the current is dependent on B. By measuring $b$ with Eqn. \ref{final}, we estimate that the theory starts to break down below $T_{\rm lower} = 0.07 K$. Thus $0.3 < T < 0.07$ K is the range where the theory should be valid (it should be emphasised that this represents a large range in the more physically relevant parameter $1/T$).  

By fitting the experimental $\lambda(B)$ to extract slope and intercept, we estimate the effective charge 
$Q_{\rm eff}(T)$, as described above. In the regime of validity, the value of $Q_{\rm eff}~$ saturates at $\sim 5 \mu_B/{\rm \AA}$, even though both slopes and intercepts depend quite strongly on temperature (Fig. \ref{fig2}). The (inverse) temperature dependence of $Q_{\rm eff}$ is illustrated in Fig. 4, where the effective charge can be seen to increase rapidly above $T_{\rm upper} = 0.3 K$, and 
to decrease slightly below $T_{\rm lower} = 0.07$ K. This direction of this slight deviation is consistent with the theory.

\section{Discussion and Conclusions}
We now compare these results with the properties, known and predicted, of Dy$_2$Ti$_2$O$_7$. As mentioned above, the predicted monopole charge for this material is $Q_{\rm theoretical} = 2 \mu/a_0 = 4.6~\mu_B/{\rm \AA} $~\cite{CMS}. Fig. \ref{fig4} illustrates how the experimentally measured charge very closely approaches this value in the temperature range $T_{\rm lower}< T < T_{\rm upper}$. The actual value of $T_{\rm upper}$ may be rationalised on the basis of our expectations for spin ice. Thus the monopole concentration should increase as $c \times e^{-2J/T}$, with $c \ll 1$ meaning that in the above temperature range it should vary between, at most, $10^{-14}$ and $10^{-4}$. This is indeed a typical degree of dissociation for a weak electrolyte that obeys Onsager's theory (i.e. screening is negligible)~\cite {Onsager,Bass}.

The success of the method is impressive. For example, regarding Fig. 2, this data may be collapsed over a significant range of temperature and field using a single parameter that is in excellent agreement with that calculated by a microscopic theory. However, even more remarkably, the method produces {\it precisely the same data collapse} (modulo the value of the charge) over a range of experimental data on a broad variety of liquid and solid electrolytes~\cite{Onsager}. Thus using Onsager's theory, we have demonstrated a perfect equivalence between electricity and magnetism. It is further a testament to Onsager's genius that his theory is not only correct and universal, but also robust, in that it may be applied to straightforward experimental measurements of only modest precision~\cite{Onsager}. As already mentioned, the robust quality of the method stems from the fact that it deals with relative rather than absolute quantities. In our case, this means that the muon spin rotation measurements are not subject to interpretation: many different analysis protocols would yield essentially the same result. 

Our $\mu$SR measurements are in full agreement with other $\mu$SR studies of Dy$_2$Ti$_2$O$_7$~\cite{Lago} as well as bulk relaxation measurements~\cite{Schiffer}. Using ultra low field magnetization measurements we have independently verified that a spin ice freezing occurs at around  $T = 0.7 K$. Below this temperature the bulk magnetization responds only extremely slowly to an applied field, but there are dynamics arising from the magnetic charges. Here we have proved the detailed origin of the dynamical response discussed in Ref. ~\cite{Lago}. 

We have thus established overwhelming evidence that deconfined magnetic charges exist in a spin ice material, that they interact by Coulomb's law and are accelerated by an applied field. Our experimental results are consistent with those discussed in Refs. \cite {CMS,Jaubert}, but more directly probe the quantities of interest in the regime of temperature and field where a dilute gas of magnetic monopoles can be interrogated. We have also measured relative changes in the magnetic conductivity (Fig. 2), and proved that monopole currents exist. This demonstration raises an interesting question. As discussed by Jaubert and Holdsworth~\cite{Jaubert}, the dynamics of spin ice monopoles are to some extent constrained by their Dirac strings, and this almost certainly precludes a direct current. It is therefore notable that diffusion currents do exist, as this would suggest that macroscopic alternating currents may also be achievable in spin ice. 

In the broader context, it has been suggested that non Ohmic transport characteristics of the Wien effect make it important in biological systems~\cite{Moore}. Therefore one might imagine that there could be novel technological applications for the magnetic equivalent. This is clearly a long way off as the temperature scales are low, but it would be interesting to establish proofs of principle in this regard, and search for other systems (perhaps micromagnetic arrays~\cite{Wang}) that show similar effects. In this context, we note the universality of the current method, and recommend it as a diagnostic tool in the study of magnetic charge transport.

\vspace{1cm}
\noindent
{\bf Acknowledgements}

It is a pleasure to thank C. Castelnovo, M. J. P. Gingras, P. C. W. Holdsworth, L. Jaubert, D. F. McMorrow and R.Moessner for useful discussions.

\vspace{1cm}

\section{Methods}
High quality single crystals of Dy$_2$Ti$_2$O$_7$ were grown using the floating zone method, and aligned using an X-ray Laue camera. In the experiment, the crystals were aligned with the initial spin polarisation of the muons along the [110] axis, with the magnetic field applied at right angles to this axis. In $\mu SR$ experiments, one measures the asymmetry of the muon beta decay as a function of time $A(t)$, which is proportional to the time evolution of the muon spin
polarization. $A(t)$ depends on the distribution of internal magnetic 
fields and their temporal fluctuations.

Our experiment was performed on the MuSR spectrometer at the ISIS facility in the UK. All data was taken upon warming after zero field cooling ($< 10 \mu T$) from 4K to 60 mK, the temperature dependence being repeated twice as a control. Transverse field spectra were fitted out to long times using $A (t) = A \exp ( - [\lambda  t] )\cos
(2\pi \upsilon t) $. The frequency of oscillations ($\upsilon$) 
can be expressed by $\upsilon=\gamma_{\mu}|B|/2\pi$, where $B$
is the average magnitude of the local field at the muon site
and $\gamma_{\mu}$ is the muon gyromagnetic ratio.  A constant, quick gaussian relaxation ($\lambda = 3\mu S^{-1}$)   was required for all the data, as expected for spin ice materials\cite{Lago}. The fitting procedure allowed the slow oscillatory relaxation $\lambda (B)$, and the applied field to vary, meaning we are directly reporting the relative change in $\lambda$. Fig. \ref{fig5} shows the time evolution of data taken at 100mK in fields of both 1 and 2 mT along with the fits to the data; the envelope of the relaxation is also shown. Background measurements have also been performed on a background silver plate to test for the field inhomogeneity in the coils: here the observed relaxation was an order of magnitude slower than in the Dy$_2$Ti$_2$O$_7$ sample. 

When zero field cooled below $0.7$ K, the susceptibility of Dy$_2$Ti$_2$O$_7$ is sufficiently small to approximate $\mu_r=1$ and to ignore demagnetizing effects. 


\begin{figure}
\includegraphics[width=0.85\linewidth]{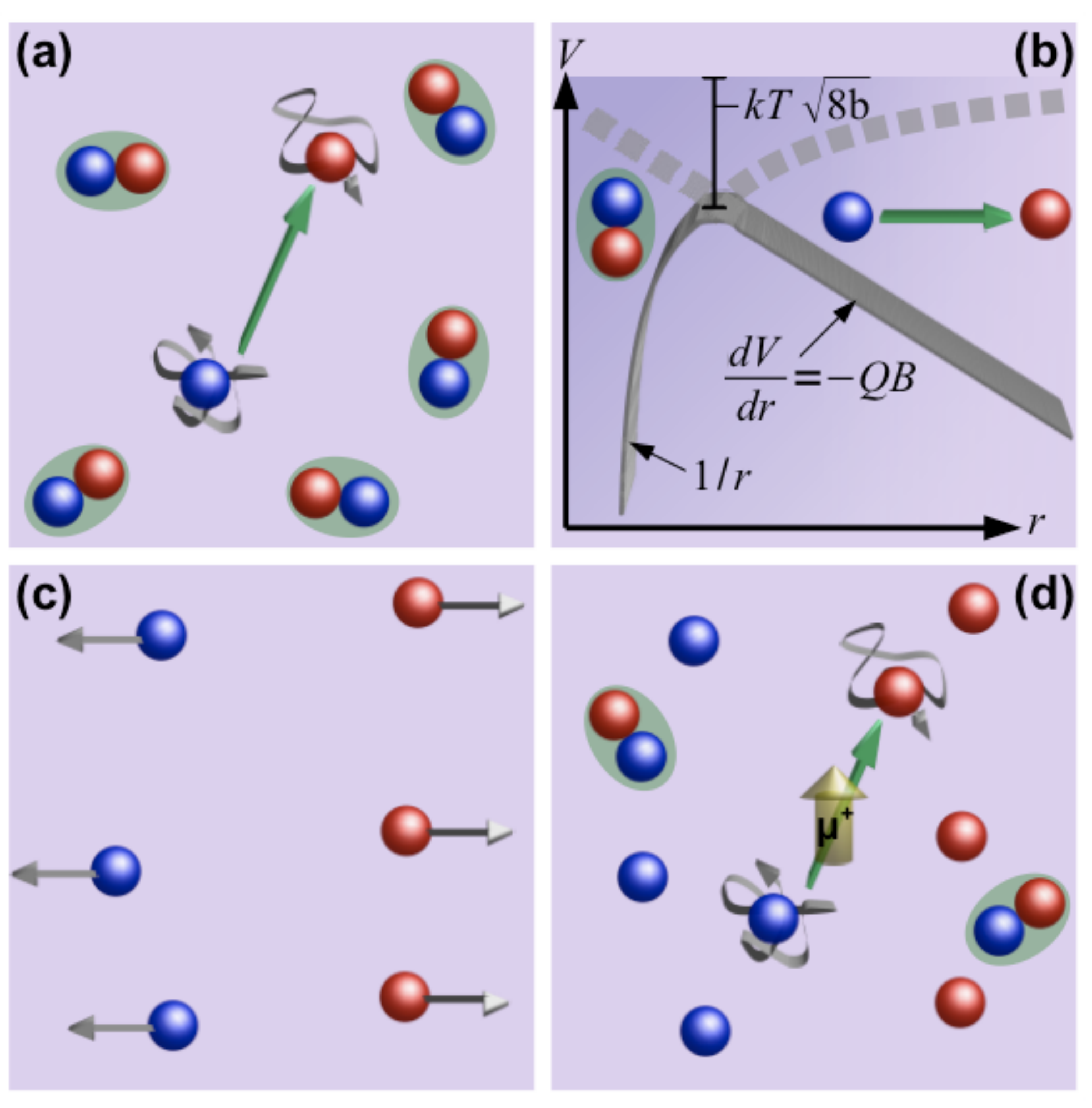}
\caption{\label{fig1} Illustration of the magnetic Wien effect, and the detection of magnetic charge by implanted muons. (a) In zero field magnetic charges occur mainly in tightly bound pairs, 
but some thermally dissociate to give a magnetic moment (green arrow) that fluctuates 
due to Brownian motion. 
(b) The competition of Coulomb potential energy ($V \sim 1/r$) and field energy $-QBr$ creates a finite energy barrier to dissociation: the stronger the field, the lower the barrier and more dissociation occurs. 
(c) On application of a transverse field,  many pairs dissociate as individual charges are accelerated by the field.
(d) After applying the field the pairs remain separated while more bound pairs form to restore equilibrium. Magnetic moment fluctuations due to free charges (green arrows) produce local fields that  are detected by an implanted muon ($\mu^{+}$).
}
\end{figure}

\vspace{1cm}
\noindent
\begin{figure}
\includegraphics[width=0.85\linewidth]{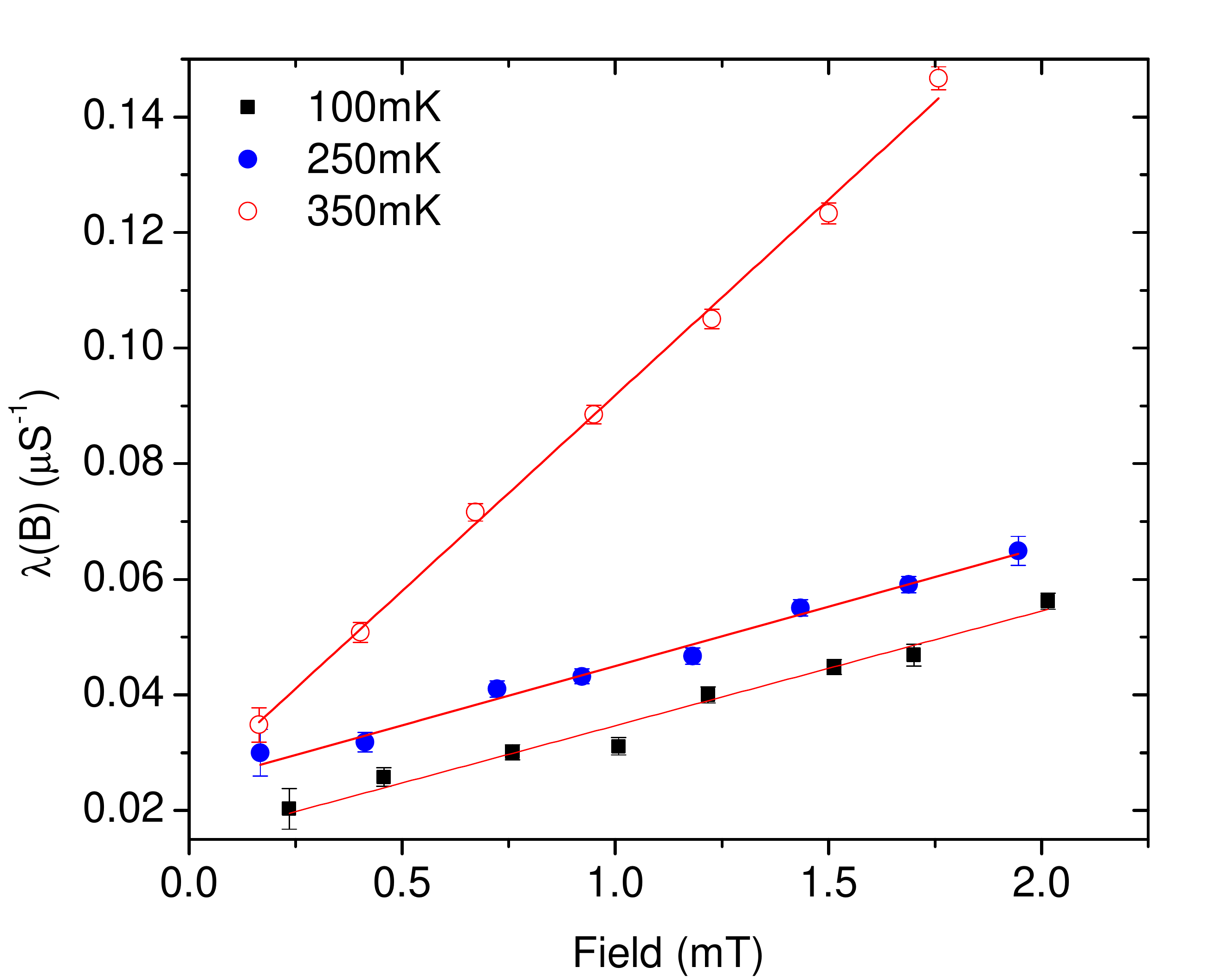}
\caption{\label{fig2} A quantity that is proportional to magnetic charge conductivity, illustrating a non-Ohmic increase in conductivity with field. The measured muon relaxation rate $\lambda (B)$ is plotted as a function of field at selected temperatures. Linear fits to the data (lines) are shown, from which it is possible to obtain the intercept, slope and consequently the effective charge $Q_{\rm eff}(T)$.}
\end{figure}

\vspace{1cm}
\noindent
\noindent
\begin{figure}
\includegraphics[width=0.85\linewidth]{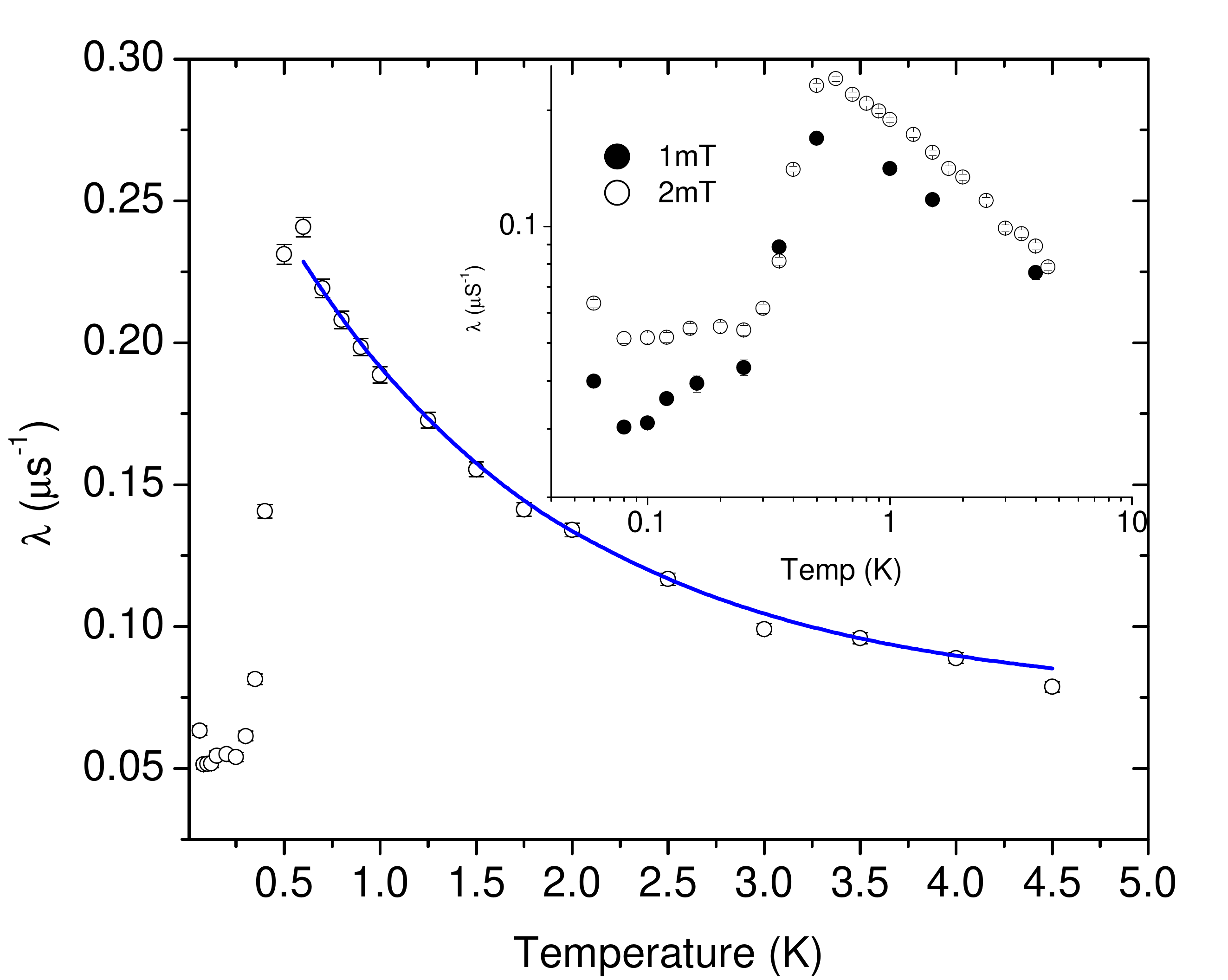}
\caption{\label{fig3} The temperature dependence of the muon relaxation rate $\lambda$ measured at an applied magnetic field of 2 mT. The high temperature regime follows the expected activated behaviour as described in the text (fit is shown as a solid line). At low temperature  $\lambda$ is proportional to the monopole concentration. Its rapid increase above $T_{\rm upper} = 0.3 K$ marks a crossover from the regime of weak screening to strong screening of the charges. The inset shows the main figure on a logarithmic scale along with data taken in 1mT. The fact that the apparently irregular behaviour at low temperature is encompassed by the theory helps demonstrate that we are indeed measuring magnetic charge transport.}
\end{figure}

\vspace{1cm}
\noindent
\begin{figure}
\includegraphics[width=0.85\linewidth]{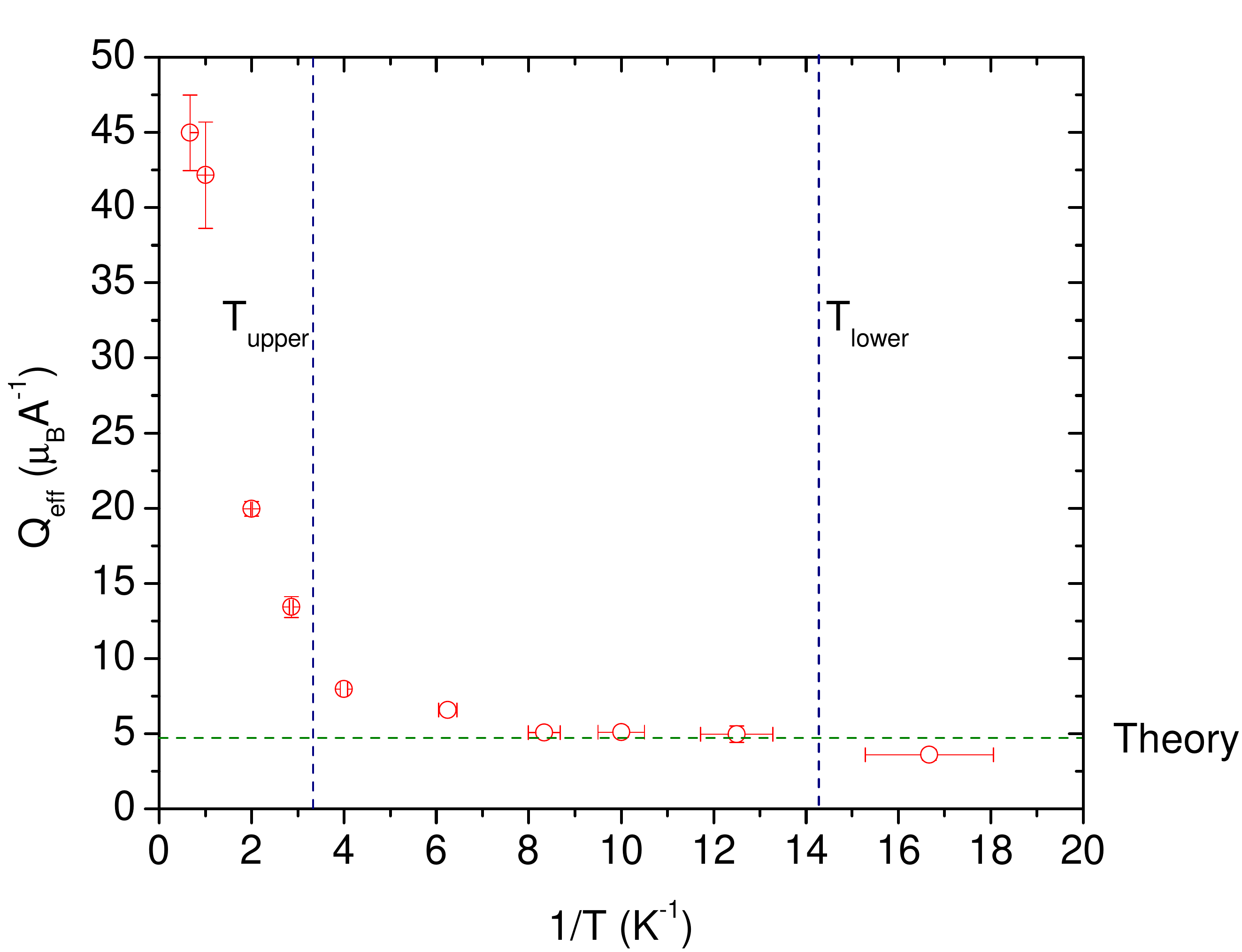}
\caption{\label{fig4}
Experimentally measured `elementary' magnetic charge $Q_{\rm eff}$ in Dy$_2$Ti$_2$O$_7$. 
Onsager's theory is valid in the regime $T_{\rm lower}< T < T_{\rm upper}$ where the magnetic charges are unscreened. The horizontal green line marks the theoretical prediction of Castelnovo, Moessner and Sondhi~\cite{CMS}.  
}
\end{figure}

\vspace{1cm}
\noindent
\begin{figure}
\includegraphics[width=0.85\linewidth]{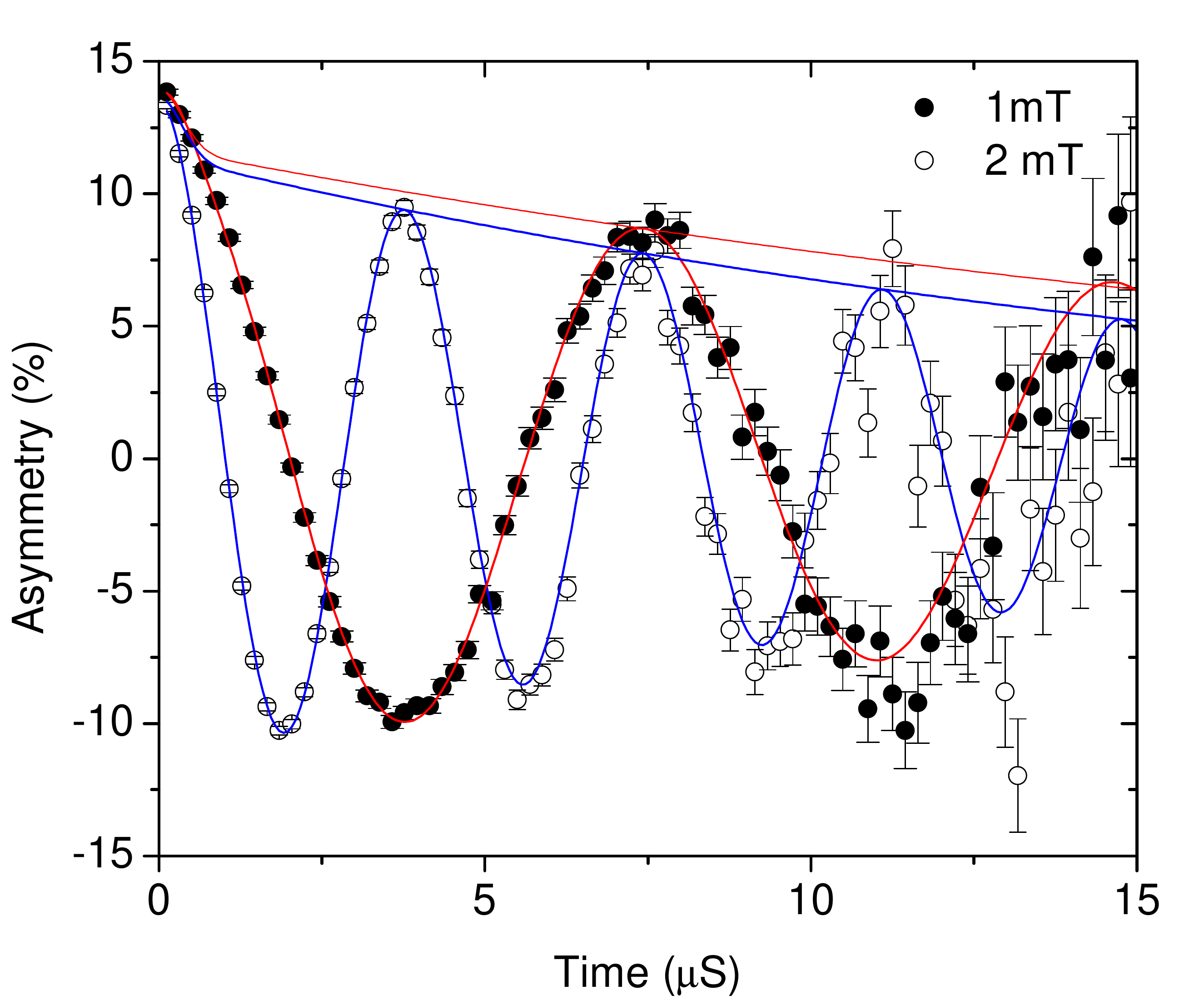}
\caption{\label{fig5} The muon relaxation as a function of time at 100 mK in an applied field of 1mT (solid circles) and 2 mT (open circles), along with the respective fits to the data. The envelope of each function is also shown, clearly showing a difference in relaxation at long time.}
\end{figure}

\end{document}